\documentclass[11pt]{article}
\usepackage{authblk}
\usepackage[T1]{fontenc}
\usepackage[utf8]{inputenc}
\usepackage[margin=1in]{geometry}
\usepackage{graphicx}
\usepackage[labelformat=empty]{caption}
\usepackage{amsmath,amssymb}
\usepackage{siunitx}
\usepackage{textcomp}
\usepackage{bm}
\usepackage{cite}
\usepackage{hyperref}
\usepackage{float}
\makeatletter

\def\@fnsymbol#1{\ensuremath{%
  \ifcase#1\or
    \dagger\or      
    \ddagger\or     
    *\or            
    \mathsection\or
    \mathparagraph\or
    \|\or
    **\or
    \dagger\dagger\or
    \ddagger\ddagger
  \else\@ctrerr\fi}}
\makeatother

\newcommand{\equal}{\textsuperscript{\textdagger}}     
\newcommand{\joint}{\textsuperscript{\textdaggerdbl}}  
\newcommand{\corr}{\textsuperscript{*}}                

\title{Self-supervised prior learning improves structured illumination microscopy resolution}

\author[1,2]{Ze-Hao Wang\equal}
\author[1,2]{Tong-Tian Weng\equal}
\author[1,2]{Xiang-Dong Chen}
\author[1,2]{Guang-Can Guo}
\author[1,2]{Fang-Wen Sun\joint\corr}
\author[3,4]{Tian-Long Chen\joint\corr}

\affil[1]{CAS Key Laboratory of Quantum Information, University of Science and Technology of China, Hefei 230026, China.}
\affil[2]{CAS Center for Excellence in Quantum Information and Quantum Physics, University of Science and Technology of China, Hefei 230026, China.}
\affil[3]{Computer Science and Artificial Intelligence Laboratory, Massachusetts Institute of Technology, Massachusetts 02139, United States.}
\affil[4]{Department of Biomedical Informatics, Harvard University, Massachusetts 02138, United States.}

\date{}

\begin{document}
\maketitle

\setcounter{footnote}{0} 
\footnotetext{These authors contributed equally to this work.}
\footnotetext{These authors jointly supervised this work.}
\footnotetext{Email: \texttt{fwsun@ustc.edu.cn}, \texttt{tianlong@mit.edu}.}

% ...

\begin{abstract}
Structured illumination microscopy (SIM) is a wide-field super-resolution technique normally limited to roughly twice the diffraction-limited resolution ($\approx 100$--$200$~nm). Surpassing this bound is a classic ill-posed inverse problem: recovering high-frequency structure from band-limited raw data. We introduce SIMFormer, a fully blind SIM reconstruction framework that learns a powerful, data-driven prior directly from raw images via self-supervision. This learned prior regularizes the solution and enables reliable extrapolation beyond the optical transfer function cutoff, yielding an effective resolution of approximately 45~nm. We validate SIMFormer on synthetic data and the BioSR dataset, where it resolves features such as flattened endoplasmic reticulum lipid bilayers previously reported to require STORM-level resolution. A self-distilled variant, SIMFormer+, further improves noise robustness while preserving high resolution at extremely low photon counts. These results show that learned priors can substantially extend SIM resolution and robustness, enabling rapid, large-scale imaging with STORM-level detail.
\end{abstract}

\noindent\textbf{One-sentence summary.}
By using a self-supervised learned prior to address the ill-posed nature of image reconstruction, SIMFormer achieves 45~nm resolution in structured illumination microscopy, revealing endoplasmic reticulum details with a clarity previously requiring STORM-level techniques.

\section*{Introduction}

In the field of super-resolution imaging, various methods such as stimulated emission depletion microscopy (STED)~\cite{Hell1994STED}, stochastic optical reconstruction microscopy (STORM)~\cite{Rust2006STORM}, and photoactivated localization microscopy (PALM)~\cite{Manley2008PALM} have greatly enhanced resolution beyond the diffraction limit. Among them, structured illumination microscopy (SIM)~\cite{Gustafsson2000SIM} is widely adopted because of its fast imaging speed, wide field of view, and relatively simple optical setup. SIM relies on computational reconstruction to extend the optical transfer function (OTF) by projecting structured light patterns onto a sample. This process encodes high-frequency sample information into detectable low-frequency moiré fringes, conventionally doubling the microscope's native resolution.

Achieving resolution beyond this twofold enhancement presents a classic ill-posed inverse problem~\cite{Tian2025CompSuperRes}. Because the raw data are band-limited by the system's OTF, there are infinitely many high-resolution structures that could produce the same observed low-resolution image. To determine a unique and physically plausible solution, one must introduce additional constraints, or prior knowledge, about the sample's structure~\cite{Tian2025CompSuperRes}. Advanced algorithms have incorporated handcrafted priors, such as sparsity and continuity, to improve resolution~\cite{Zhao2022SparseSIM}, but these priors can be overly simplistic and sensitive to parameter tuning, limiting their generalizability in practice~\cite{Chen2023SIMReview}.

The deep learning revolution inspired a shift toward data-driven learnable priors, utilizing deep neural networks trained with labeled data for SIM reconstruction. Notable supervised approaches include U-Net-SIM~\cite{Jin2020UNetSIM}, DFCAN~\cite{Qiao2021DFCAN}, and DFGAN~\cite{Qiao2021DFCAN}, employ supervised learning approaches. These methods rely on high-resolution images—either simulated or experimentally measured with other microscopy techniques—to serve as ground truth for training. However, the practical application of such supervised methods is highly limited by their poor generalization ability when applied to new, unseen testing samples~\cite{Belthangady2019DLReview,Weigert2018CARE}. This lack of adaptability stems from out-of-distribution generalization issues, where discrepancies between the labeled training data and new experimental data undermine the reliability of the results.

Here we introduce SIMFormer, a Transformer-based~\cite{Vaswani2017Attention,Dosovitskiy2021ViT} self-supervised learning framework that resolves the ill-posed challenge of SIM reconstruction by learning a powerful, data-driven prior directly from raw microscopy data. Using a masked autoencoder (MAE)~\cite{He2022MAE,Feichtenhofer2022STMAE} training strategy, SIMFormer is compelled to learn the intrinsic statistical patterns of biological structures. This learned prior acts as a sophisticated, adaptive regularizer, guiding the reconstruction to solutions that are not only consistent with the measured low-frequency data but also conform to the learned characteristics of plausible biological structures.

Such a process can be viewed as reliable spectral extrapolation, in which the network leverages the available low-frequency information and the strong data-driven prior to infer the most probable high-frequency information that was not directly measured. Consequently, SIMFormer achieves a resolution of approximately 45~nm---nearly a fivefold improvement over the diffraction limit. The framework is fully blind, simultaneously estimating the emitters, illumination patterns, background, and point spread function (PSF) from the raw data alone. To further enhance noise robustness under low-photon-count conditions, we introduce SIMFormer+, a refined model developed through self-distillation~\cite{Zhang2019SelfDistill}. The effectiveness of SIMFormer is thoroughly validated through extensive evaluations on both synthetic and real-world SIM datasets, including the BioSR benchmark~\cite{Qiao2021DFCAN}, which comprises diverse biological structures such as clathrin-coated pits (CCPs), endoplasmic reticula (ER), microtubules (MTs), and F-actin filaments. Together, these results highlight the broad applicability and practical utility of SIMFormer for super-resolution fluorescence imaging.

\section*{Results}

\subsection*{Prior learning via low-dimensional subspace projection in SIMFormer}

Conventional maximum a posteriori (MAP) image reconstruction methods address ill-posedness by incorporating explicit, handcrafted priors such as sparsity and smoothness to regularize the solution~\cite{Farsiu2004SR,Bouman1993GGMRF,Dey2006RLTV}. In contrast, SIMFormer utilizes learnable priors that emerge from the data itself through the use of MAE~\cite{He2022MAE}. By reconstructing images from aggressively masked inputs, often discarding around 75\% of patches during training, MAE is compelled to capture only the most predictive feature components from the visible patches~\cite{Zhang2022MaskMatters}. This masking-driven process confines the representation to a task-driven, data-adaptive, low-dimensional subspace, effectively creating an implicit prior learned directly from the training data without requiring any manual assumptions on emitters~\cite{Zhang2022MaskMatters}. One can formulate the MAE objective in MAP terms as minimizing a reconstruction error with a learned regularizer: $\min_{z} \|D(z) - x\|_2^2 + \lambda R(z)$, where $D(z)$ reconstructs the target image $x$ and $R(z)$ encodes the learned prior on the latent code.

\subsection*{Comparison with existing methods}

SIMFormer's approach to generating and utilizing priors fundamentally distinguishes it from other SIM reconstruction methods. Traditional algorithms such as Sparse-SIM~\cite{Zhao2022SparseSIM} and Blind-SIM~\cite{Mudry2012SpeckleSIM} depend on iterative optimization guided by explicit, handcrafted priors and parametric assumptions about the illumination pattern. These non-learning methods are often computationally intensive and sensitive to parameter tuning~\cite{Chen2023SIMReview}. Recent deep learning approaches operate on varied principles. Some, such as the Swin Transformer-based denoising method by Shah \emph{et al}.~\cite{Shah2024SwinSIMDenoise} and the self-supervised multimodal denoising method by Chen \emph{et al}.~\cite{Chen2024SelfSupDenoiseSIM}, are post-process denoisers applied to previously reconstructed images; consequently, they cannot recover spatial frequencies lost during initial reconstruction. Others, for instance, the physics-informed network by Burns and Liu~\cite{Burns2023PINNSIM}, perform full reconstruction but utilize an untrained network’s implicit prior. These methods require known illumination patterns and are bound by the theoretical resolution limit. SIMFormer, in contrast, is a full, end-to-end blind reconstruction engine that leverages a strong, data-driven prior learned through self-supervision. This learned prior functions as an adaptive regularizer, allowing for reliable spectral extrapolation beyond conventional resolution limits, a capability not provided by the principles of other methods.

\subsection*{The network architecture of SIMFormer}

SIMFormer introduces a novel architecture for fully blind SIM reconstruction which contains four components. (1) Data processor. It takes raw SIM images as an input stack consisting of the three phases from each of the SIM image's three directions. The input stack is divided into $3\times 16\times 16$ 3D cubes, and then they are transformed into 768-dimensional vectors by patch embedding. (2) Encoder. The embedded data is further processed by a transformer encoder derived from ViT~\cite{Dosovitskiy2021ViT} (ViT-base) with 12 transformer layers. Meanwhile, an adapter is introduced into the encoder to accelerate training and effectively improve dense prediction capabilities~\cite{Chen2023ViTAdapter}. (3) Decoder. The encoded features are processed by a decoder consisting of 12 transformer layers, the output of which is passed through three convolutional decoding heads to separately output emitters, light patterns, and background to reconstruct the raw SIM images. Particularly, for the light pattern, a low-rank coding module is introduced before the convolutional decoder head to progressively compress the rank of the light pattern representation during training. (4) PSF. It is modeled by feeding a fixed random embedding to a learnable convolutional neural network. Overall, the raw SIM images are synthesized using the emitters, light patterns, background, and PSF following the principles of SIM imaging. See Extended Data Fig.~1 for details.

\subsection*{Multi-stage self-supervised training}

Our SIMFormer adopts a multi-stage self-supervised training approach inspired by curriculum learning~\cite{Bengio2009Curriculum}. At each stage, the masking rate and the compression ratio of the low-rank coding will be adjusted. Specifically, the masking rate decreases, while the compression ratio gradually increases, following a multi-stage schedule. This design aims to prevent the model from converging to trivial PSF solutions in the early stages and promote a stable progression toward low-rank coding for the light pattern.

Our training objective involves four parts as follows: (1) a reconstruction loss function to measure the difference between the synthesized and raw SIM images; (2) a total variation loss function to enforce smoothness in the predicted light pattern; and (3) a total variation loss function for PSF and light pattern smoothness, along with (4) a mean absolute error loss to ensure alignment between the PSF centroid and the image center. See Methods for further details.

\subsection*{Self Distillation}

Knowledge distillation~\cite{Hinton2015Distilling} is a machine learning technique where a student model is trained to replicate the behavior of a teacher model. In this study, we adopt a self-distillation~\cite{Zhang2019SelfDistill,Zhang2022SelfDistillationSurvey,Pham2022RevisitingSelfDistill} approach, where the student model shares the same architecture as the teacher. By using the teacher's predictions as soft labels, the student model learns to refine and enhance the original outputs, resulting in improved accuracy and stronger noise robustness. This refined model is referred to as SIMFormer+.

\subsection*{Data preparation}

We demonstrate SIMFormer’s capabilities using both synthetic and real-world data. Several datasets with varying fluorophore structures, light pattern spatial periods, noise levels, line sparsity, tube widths, and ring radii are generated for synthetic validation. In these datasets, the fluorescence wavelength is set to 488 nm, the objective's numerical aperture (NA) is 1.3, and the pixel size is 62.6 nm. The PSF is simulated using the Debye integral method~\cite{Zhang2007GaussianPSF}, corresponding to a diffraction limit of 229 nm. The light pattern is modeled as a cosine function, oriented in three directions with three phases for each direction. Using the emitters, light patterns, and the PSF, we synthesized raw SIM images and subsequently reconstructed super-resolution images using conventional algorithm (Supplementary Note 2 and 3). For each parameter configuration, we generated 100 samples for the training set and another 100 samples for the test set. As for real-world validation, the BioSR dataset is adopted, which contains raw SIM images of various biological structures, including CCPs (54 cells), ER (68 cells), MTs (55 cells), and F-actin (51 cells), along with the corresponding super-resolution images obtained through the conventional SIM reconstruction algorithm.

\subsection*{Comparative Evaluation and Metrics}

We use both synthetic and real experimental data to evaluate the performance of SIMFormer. Meanwhile, it is compared with representative existing approaches including conventional SIM reconstruction, Sparse-SIM~\cite{Zhao2022SparseSIM}, and Blind-SIM~\cite{Mudry2012SpeckleSIM}. Sparse-SIM is currently the state-of-the-art SIM reconstruction algorithm with the highest resolution~\cite{Chen2023SIMReview}, while Blind-SIM is a widely used blind reconstruction algorithm~\cite{Lee2021MetamaterialNanoscopy}. Using the labeled synthetic dataset, we examine the superiority of  SIMFormer over Sparse-SIM and Blind-SIM under two challenging conditions: (1) SIMFormer is trained and makes predictions in a fully blind manner, whereas Sparse-SIM has access to experimental parameters, and Blind-SIM uses the true PSF; (2) The performance of Sparse-SIM and Blind-SIM is fine-tuned using genetic algorithms to achieve optimal results (Supplementary Note 4). However, SIMFormer has not been optimized due to computational resource limitations. In real-world experiments, Blind-SIM used the same parameters as in the synthetic data experiments, while Sparse-SIM employed new parameters to achieve better performance (Supplementary Note Table 2).

The synthetic dataset provides actual emitters as ground truth, allowing us to use the normalized root-mean-square error (NRMSE) and the multiscale structural similarity index (MS-SSIM) to evaluate the reconstructed images. NRMSE quantifies the overall difference in intensity between the predicted and actual images, providing a measure of the error magnitude. In contrast, MS-SSIM assesses the structural similarity between the images at multiple scales, focusing on perceived visual quality related to structural information. Note that we use NRMSE with added Gaussian blur to measure the spatial distance error of the emitters (Supplementary Note 5). Since the resolution achieved by Sparse-SIM and SIMFormer surpasses that of conventional SIM reconstruction in the BioSR dataset (so we don't have ground truth), we focus on the discriminability of regions of interest and compare the results with prior knowledge from STORM imaging of ERs~\cite{Wang2022ERTubules}. Additionally, we apply decorrelation analysis~\cite{Descloux2019Decorrelation} to measure resolution (Supplementary Note 6). For the noise robustness evaluation, we use fairSIM~\cite{Muller2016fairSIM} to generate conventional SIM reconstruction results under high noise levels, as these results are not available in BioSR.

\subsection*{SIMFormer demonstrates highly accurate super-resolution imaging on synthetic data}

We first evaluate SIMFormer on synthetic filament structures, where fluorophores are simulated using randomly generated curves based on Bessel functions (Supplementary Note 2). As shown in Fig.~\ref{fig:fig2}d, the NRMSE results are as follows: Widefield 1.15, conventional SIM 0.87, Blind-SIM 0.67, Sparse-SIM 0.59, SIMFormer 0.32, and SIMFormer+ 0.30. Notably, SIMFormer+ achieves a 49.5\% improvement in NRMSE over Sparse-SIM and a 55.7\% improvement over Blind-SIM. For MS-SSIM, the values are: Widefield 0.64, conventional SIM 0.63, Blind-SIM 0.75, Sparse-SIM 0.85, SIMFormer 0.90, and SIMFormer+ 0.92. Here, SIMFormer+ demonstrates an 8.2\% improvement over Sparse-SIM and a 22.3\% improvement over Blind-SIM. As shown in Fig.~\ref{fig:fig2}g, the resolution values from decorrelation analysis indicate: Widefield 219.4 nm, conventional SIM 122.0 nm, Blind-SIM 67.3 nm, Sparse-SIM 69.2 nm, SIMFormer 48.5 nm, and SIMFormer+ 49.3 nm. SIMFormer+ achieves approximately 2.5 times the resolution of conventional SIM and 1.4 times better resolution than both Blind-SIM and Sparse-SIM.

A closer examination of the zoomed-in regions, particularly the 46 nm buckle structure, highlights several important observations that support the quantitative results: (1) conventional SIM reconstruction fails to resolve the buckle structure, (2) Blind-SIM reconstruction implies a rough structure but introduces significant artifacts, (3) Sparse-SIM shows fewer artifacts, yet still cannot distinctly resolve the structure, (4) SIMFormer produce a rough buckle structure with fewer reconstruction artifacts, attributed to the learned priors, and (5) SIMFormer+ produce a clear buckle structure compared to its vanilla counterpart, thanks to our self-distillation designs which brings more reconstruction robustness. Moreover, Supplementary Fig. 1 presents the reconstruction results for samples varying from sparse to dense, where SIMFormer and SIMFormer+ show a significant performance advantage, particularly in dense samples. 

Next, we evaluate SIMFormer on synthetic microtubule structures. As shown in Fig.~\ref{fig:fig2}e, the NRMSE results are: Widefield 1.16, conventional SIM 0.83, Blind-SIM 0.70, Sparse-SIM 0.61, SIMFormer 0.34, and SIMFormer+ 0.33. SIMFormer+ demonstrates a 47.0\% improvement over Sparse-SIM and a 53.2\% improvement over Blind-SIM. For MS-SSIM, the values are: Widefield 0.56, conventional SIM 0.59, Blind-SIM 0.74, Sparse-SIM 0.82, SIMFormer 0.86, and SIMFormer+ 0.91, where SIMFormer+ achieves a 10.5\% improvement over Sparse-SIM and a 22.2\% improvement over Blind-SIM. As shown in Fig.~\ref{fig:fig2}h, resolution values from decorrelation analysis indicate: Widefield 238.3 nm, conventional SIM 127.8 nm, Blind-SIM 66.3 nm, Sparse-SIM 73.3 nm, SIMFormer 55.3 nm, and SIMFormer+ 51.7 nm. SIMFormer+ achieves approximately 2.5 times the resolution of conventional SIM, 1.3 times better resolution than Blind-SIM, and 1.4 times better resolution than Sparse-SIM.

In Fig.~\ref{fig:fig2}b, the zoomed-in regions highlight microtubule structures that are 65 nm wide. These structures are not clearly resolved by any method except SIMFormer and SIMFormer+, as they exceed the recognition limits of blind reconstruction and hand-crafted priors, leading to ambiguity. Extended Data Fig.~2 shows that while both Sparse-SIM and Blind-SIM can resolve wider microtubule structures (approximately 126 nm), Sparse-SIM performs relatively better than Blind-SIM.

Finally, we simulate ring-shaped fluorophores with varying radii to validate SIMFormer on ring structures. As shown in Fig.~\ref{fig:fig2}f, the NRMSE results are: Widefield 1.19, conventional SIM 0.85, Blind-SIM 0.65, Sparse-SIM 0.65, SIMFormer 0.32, and SIMFormer+ 0.29. SIMFormer+ improves NRMSE by 55.9\% compared to Sparse-SIM and Blind-SIM. For MS-SSIM, the values are: Widefield 0.68, conventional SIM 0.70, Blind-SIM 0.85, Sparse-SIM 0.88, SIMFormer 0.91, and SIMFormer+ 0.95, showing a 7.9\% improvement over Sparse-SIM and 10.8\% over Blind-SIM. Furthermore, resolution values from decorrelation analysis in Fig.~\ref{fig:fig2}i indicate: Widefield 238.7 nm, conventional SIM 135.3 nm, Blind-SIM 71.9 nm, Sparse-SIM 72.3 nm, SIMFormer 48.7 nm, and SIMFormer+ 47.9 nm. SIMFormer+ achieves 2.8 times the resolution of conventional SIM and 1.5 times better resolution than both Blind-SIM and Sparse-SIM. In Fig.~\ref{fig:fig2}c, SIMFormer and SIMFormer+ successfully resolve ring structures about 75 nm in diameter, which conventional SIM, Sparse-SIM, and Blind-SIM do not. Extended Data Fig.~3 shows Sparse-SIM resolving rings 125 nm or larger, while Blind-SIM resolves rings 165 nm or larger.

Overall, considering discernibility, accuracy, and resolution, the ranking of methods is SIMFormer+, SIMFormer, Sparse-SIM, Blind-SIM, and conventional SIM. Quantitative and visual metrics consistently support this ranking assessment across all benchmarks. These performance differences highlight a fundamental distinction in how each method addresses the ill-posed nature of SIM reconstruction. Methods like Sparse-SIM impose hand-crafted priors, such as sparsity, which can be overly rigid for complex structures. In contrast, SIMFormer utilizes a flexible and powerful prior learned directly from data. This self-supervised approach forces the model to understand the underlying statistical patterns of microscopy images, thereby acting as an adaptive regularizer. Consequently, when faced with ambiguous regions, such as the dense 46 nm buckle, SIMFormer resolves the structure more reliably, avoiding the tendency of methods like Sparse-SIM to interpret such areas as fine lines.
\subsection*{SIMFormer achieves 45 nm resolution on the BioSR dataset}

Here, we use the BioSR dataset to evaluate SIMFormer in real-world conditions. The BioSR dataset consists of four structure types—CCPs, ER, MTs, and F-actin filaments—ranging from sparse to dense. With a fluorescence wavelength of 488 nm and an NA of 1.3, the diffraction limit is 229 nm. We process the raw SIM images using Blind-SIM, Sparse-SIM, and SIMFormer, alongside conventional SIM reconstructions provided in the BioSR dataset. Then, self-distillation is performed on the SIMFormer results to obtain SIMFormer+. As shown in Fig.~\ref{fig:fig3}, these results are compared side-by-side with the wide-field imaging results.

CCPs are specialized regions of the cell membrane responsible for endocytosis, where cells internalize essential molecules such as nutrients and receptors. Structurally, CCPs are ring-like assemblies of the protein clathrin, typically measuring 50 to 150 nm in diameter, and their size is crucial for vesicle formation and cargo selection~\cite{McMahon2011Clathrin}. Fig.~\ref{fig:fig3}a presents the comparison results for CCPs. In previous synthetic data experiments, SIMFormer demonstrated its capability to resolve 75 nm ring structures. In this case, only SIMFormer can detect the 78 nm micro-ring, while all methods identify the 125 nm micro-ring. SIMFormer+ achieves consistent results with SIMFormer. Additional results are shown in Supplementary Fig. 2.

The ER is a key organelle involved in protein synthesis, lipid metabolism, and calcium storage, forming a tubular network with nanoscale tubules around 50--100 nm in diameter, as confirmed by STORM microscopy~\cite{Wang2022ERTubules}. These tubules play a critical role in material distribution and maintaining cellular homeostasis. Fig.~\ref{fig:fig3}b presents a side-by-side comparison of ER images, showing that while Sparse-SIM provides sharper and more coherent results than BioSR SIM, it fails to resolve the tubular structures. Blind-SIM shows some microtubules but introduces many artifacts. In contrast, SIMFormer clearly reveals the 80 nm width microtubules closely matching STORM results. SIMFormer+ achieves similar results while improving robustness and suppressing artifacts in low-light regions. More results are shown in Extended Data Fig.~4.

MTs are essential components of the cytoskeleton, providing structural support, facilitating intracellular transport, and playing key roles in cell division and signaling. These dynamic, tube-like polymers, typically 25 nm in diameter, are composed of tubulin proteins and are critical for maintaining cell shape and organizing organelles~\cite{Desai1997MTDynamics}. Fig.~\ref{fig:fig3}c shows the comparison results for MTs, with all methods resolving the filamentous structures. However, SIMFormer offers clearer results in regions where the microtubules overlap, particularly in the zoomed-in sections highlighting intersecting structures. Additionally, SIMFormer+ further reduces artifacts. Extended Data Fig.~5 demonstrates that SIMFormer can resolve microtubule structures as narrow as 34 nm in width.

F-actin (filamentous actin) is a critical component of the cytoskeleton, playing a vital role in cellular processes such as movement, shape maintenance, and division. These thin, flexible filaments, about 7 nm in diameter, are made up of actin subunits and form dense networks that provide structural support and facilitate cellular motility~\cite{Cooper2022CellBook}. Fig.~\ref{fig:fig3}d presents the comparison results for F-actin filaments. SIMFormer resolves these dense filamentous structures more clearly than other methods, while SIMFormer+ further reduces artifacts in low-light regions. Additional results are shown in Supplementary Fig. 3.

Fig.~\ref{fig:fig3}e--h shows the resolution calculated using decorrelation analysis for BioSR SIM, Sparse-SIM, Blind-SIM, SIMFormer, and SIMFormer+ across the ER, CCPs, MTs, and F-actin filaments datasets. SIMFormer and SIMFormer+ demonstrate superior resolution compared to the other methods, reaching on average 46.5 nm and 47.6 nm (4.9 and 4.8 times the diffraction limit), consistent with their previously observed resolving capabilities in the synthetic data experiment. SIMFormer+ shows minimal loss in resolution while significantly improving robustness and suppressing artifacts. In contrast, conventional SIM reconstruction (BioSR SIM) achieves 1.7 times the diffraction limit, Sparse-SIM reaches 2.3 times, and Blind-SIM, while achieving 3.1 times the diffraction limit, suffers from significant artifacts. 

The practical power of SIMFormer’s learned prior is clearly demonstrated by its ability to resolve 80 nm ER tubules, a level of detail previously associated with STORM imaging~\cite{Wang2022ERTubules}. This achievement is not merely image enhancement; it represents a form of reliable spectral extrapolation beyond the conventional 2x resolution limit. Critically, the learned prior overcomes the limitations of hand-crafted rules, such as sparsity, which are ill-suited for complex biological structures like the interconnected ER network. Reliance on such rigid priors can cause algorithms like Sparse-SIM to misinterpret these dense tubular structures as simple fine lines. In contrast, SIMFormer’s adaptive, data-driven prior allows it to plausibly infer high-frequency information that is inaccessible to traditional methods, thereby breaking the conventional resolution barrier to reveal these ultrastructures with higher fidelity.

\subsection*{SIMFormer predicts accurate light patterns with self-supervised learned priors}

Trained by reconstructing raw SIM images, the accuracy of SIMFormer’s light pattern prediction is directly tied to the accuracy of its emitter predictions. We evaluate SIMFormer’s light pattern prediction on synthetic datasets and compare it to Blind-SIM. As shown in Fig.~\ref{fig:fig4}d, SIMFormer achieves an average NRMSE of 0.39, while Blind-SIM records 0.51. Notably, the light pattern evaluation is performed only in regions near the emitters (Supplementary Note 2), as the reconstruction of the light pattern in areas far from the emitters has no impact on the emitter reconstruction. As shown in Fig.~\ref{fig:fig4}a--c, due to the learned priors and low-rank coding, SIMFormer’s predicted light patterns are significantly closer to the true ones, supporting the quantitative comparison. Fig.~\ref{fig:fig4}e--h shows light pattern predictions for CCPs, ERs, MTs, and F-actin from the BioSR dataset. SIMFormer accurately predicts the cosine light pattern across all structures, whereas Blind-SIM produces anomalous light patterns. 

\subsection*{SIMFormer achieves both high resolution and strong noise robustness through self-distillation}
 We validate the noise robustness of SIMFormer and SIMFormer+ using both synthetic and BioSR datasets. In the synthetic datasets, we test the three previously mentioned structures, with a light pattern spatial period of 313 nm and average photon numbers set to 0.5, 1, 10, 100, 500, and 1000, respectively. The quantitative results are shown in Fig.~\ref{fig:fig5}b. NRMSE and MS-SSIM measurements across varying photon numbers indicate that SIMFormer and SIMFormer+ comprehensively outperform all other methods. A detailed analysis of the NRMSE values reveals that SIMFormer+ significantly enhances SIMFormer’s performance under extremely low photon counts (0.5), with a 30.1\% reduction in NRMSE for synthetic filament structures, 4.8\% for synthetic microtubule structures, and 17.5\% for synthetic ring structures. As shown in Supplementary Fig. 4, SIMFormer+ maintains consistent resolution at low photon counts, while SIMFormer struggles due to artifacts that prevent decorrelation analysis. The observations in Fig.~\ref{fig:fig5}b further support these findings, demonstrating that SIMFormer+ greatly improves SIMFormer's robustness under low photon counts. Additional results are provided in Supplementary Figs. 5--7.

In Fig.~\ref{fig:fig6}, we present a side-by-side comparison of conventional SIM reconstruction, SIMFormer, and SIMFormer+ on the BioSR dataset under both low and high photon counts. To ensure that SIMFormer does not memorize low-noise data during the noise robustness evaluation, we use the first 35 cells as the training set and the remaining cells as the test set. Key conclusions are as follows: (1) SIMFormer and SIMFormer+ consistently produce fewer artifacts and greater robustness under low photon counts across all tasks, (2) SIMFormer+ generates fewer artifacts than SIMFormer under low photon counts, and (3) SIMFormer+ enhances the discernibility of biological structures, such as the ring-shaped structures in CCPs and the tubular structures in ER, compared to SIMFormer under low photon counts. Further results can be found in Extended Data Figs.~6--9. These findings suggest that the learnable priors in SIMFormer, combined with the self-distillation in SIMFormer+, deliver a significant improvement in both resolution and noise robustness.

\section*{Discussion}

In this work, we present SIMFormer, a fully blind reconstruction framework that simultaneously estimates the sample structure, PSF, light patterns, and background from raw SIM images. By integrating a learnable prior acquired through self-supervised learning with a low-rank encoding of the light pattern, SIMFormer achieves a resolution approaching 45 nm---nearly five times the diffraction limit---while significantly improving the accuracy of light pattern estimation over previous blind methods. Furthermore, with the introduction of a self-distillation step, the resulting SIMFormer+ model enhances both reconstruction accuracy and noise robustness without compromising this high resolution.

The remarkable performance of SIMFormer stems from its novel approach to learning physics-informed priors. Instead of relying on handcrafted rules, the model’s architecture is designed to follow SIM imaging principles, while its self-supervised training on raw data compels it to capture the intrinsic statistical distribution of complex biological structures. This synergy creates a powerful, data-driven regularizer that effectively constrains the solution space for the ill-posed inverse problem, thereby reducing reconstruction ambiguity. This learned prior is also modular and could be combined with traditional handcrafted priors, such as sparsity or continuity, to potentially achieve further performance gains.

SIMFormer’s design principles also naturally suggest routes for extending data-driven priors to more complex imaging scenarios, without relying on additional calibration data. By tightly coupling physical image formation models with flexible learned priors, the framework provides a template for future algorithms that must balance robustness, fidelity, and practicality in demanding experimental settings.

Furthermore, the framework can be naturally adapted to process volumetric data by using 3D patches and adding axial positional encodings. Such an extension holds the promise of achieving isotropic 3D super-resolution, which, combined with the method's inherent noise robustness, would be particularly powerful for gentle, long-term 4D live-cell imaging of dynamic processes without causing significant phototoxicity.

Beyond these practical applications and experimental paradigms, the relationship between SIMFormer’s reconstruction error and the Cramér--Rao lower bound deserves further investigation to understand the fundamental limits of this approach. Ultimately, the robust and adaptable architecture of SIMFormer shows great potential not only for super-resolution imaging but also for a wide range of other scientific analysis problems where ill-posed inverse problems are a central challenge.

\section*{Materials and Methods}

\subsection*{Encoder and Decoder}

As shown in Extended Data Fig.~1, our encoder is designed to process raw images from SIM and is inspired by Spatiotemporal Masked Autoencoders~\cite{Feichtenhofer2022STMAE}. We develop a ViT encoder capable of processing multi-frame SIM images with pre-trained ViT weights (CLIP~\cite{Radford2021CLIP}). To speed up training and improve the extraction of fine-grained features, we incorporated ViT-adapter~\cite{Chen2023ViTAdapter}. Specifically, our encoder, modified from the ViT-based architecture, converts multi-frame SIM images into multiple 768-dimensional vectors using three-dimensional patch embedding. Simultaneously, the original multi-frame SIM images are processed through a convolutional neural network-based (CNN-based) stem, resulting in an equal number of 768-dimensional vectors for each patch. Both sets are subjected to the same random masking, which masks a portion of the patches. The unmasked patches are combined with learnable position embeddings and frame position embeddings, shuffled, and fed separately into the encoder and adapter layers. These elements interact and merge during forward propagation, facilitating information exchange. The output features of the encoder are then used. A perceptron layer transforms the features of each patch into 512-dimensional vectors. A learnable token is introduced to replace the masked patches. These features are then reorganized into their original order and added with learnable positional embeddings. After this reordering, the features are fed into a decoder consisting of eight transform layers. The output of the decoder is then used by three separate prediction heads.

\subsection*{Prediction modules for emitters and background}

SIMFormer uses a simple CNN-based prediction module. This module consists of four blocks. Each block is for two-fold upsampling. Within each block, the module performs a double upsampling step, followed by three convolutional operations with filter sizes of $3\times 3$. After each convolution, a gaussian error linear unit activation function is used as the activation function. The prediction module starts with 512 channels and methodically reduces the number of channels in the output of these convolutional layers, halving it in each block. The final stage is a convolutional layer with a $1\times 1$ filter that produces a single channel output.

The emitter prediction subnetwork consists of the prediction module mentioned above, followed by a soft plus activation layer, to achieve non-zero outputs. The background prediction subnetwork incorporates the aforementioned prediction module, coupled with an average pooling layer that features a kernel size of 64. This design assumes that the background exhibits slow intensity variations across the spatial domain. 

\subsection*{Light pattern reconstruction via low-rank coding}

In SIM, the light pattern consists of only a few modes. For instance, in a typical SIM experiment, the cosine light pattern operates in three directions with nine phases, corresponding to nine distinct modes. This implies that the number of predicted light pattern modes can be reduced. To integrate this prior knowledge into SIMFormer, we employ a low-rank encoding method via an average pooling layer. Specifically, the decoder's feature output first passes through a Transformer layer, followed by a one-dimensional average pooling operation applied to each feature vector, and then through another Transformer layer before producing the final output.

In predicting the light pattern, a common assumption is that the sum of all light patterns in a set of SIM experiments yields a uniform illumination pattern~\cite{Mudry2012SpeckleSIM,Lee2021MetamaterialNanoscopy}. To ensure that the sum equals one and that the intensity of the light patterns remains non-negative, a softmax operation is applied to the output of the prediction module along the channel dimension.

\subsection*{PSF Prediction}

The PSF prediction module is a convolutional neural network that uses a fixed random tensor as input to generate the PSF. Based on the principles of Deep Image Prior~\cite{Ulyanov2018DeepImagePrior}, the neural network reframes the challenge of estimating the PSF as an optimization task within the neural network. This methodology has been demonstrated to learn the PSF directly from raw microscopy images~\cite{Wang2024ImagingMechanism}.

\subsection*{Loss Function}

We designed several functions and combined them to form a loss function to learn in a self-supervised manner from raw SIM images. These functions can be divided into two categories. The first category measures the difference between the reconstructed SIM images and the raw SIM images. In this category, we introduce the $L_1$ distance and multi-scale structural similarity (MS-SSIM). The weighted average of $L_1$ distance and MS-SSIM has been proven to be very effective in image restoration tasks~\cite{Zhao2016LossFunctions}.

In the second category, considering that the intensity-weighted center of the PSF should coincide with the geometric center of the PSF tensor to prevent drift in the prediction results, we designed a center distance function. Additionally, based on optical principles, the intensity of the PSF and light pattern should have spatial continuity, so we introduced a total variation (TV) loss for the PSF and light pattern. The loss function used in the training of SIMFormer is the weighted average of the above functions, as shown below:
\begin{align}
L_1 &= \bigl|\mathrm{SIM}_{\mathrm{pred}} - \mathrm{SIM}_{\mathrm{raw}}\bigr|,\\
L_{\mathrm{MS\text{-}SSIM}} &= 1 - \mathrm{MS\text{-}SSIM}\bigl(\mathrm{SIM}_{\mathrm{pred}}, \mathrm{SIM}_{\mathrm{raw}}\bigr),\\
\mathrm{Central} &= \left|\frac{\sum_{x,y} (x,y)\,\mathrm{PSF}(x,y)}{\sum_{x,y} \mathrm{PSF}(x,y)} - (c_x,c_y)\right|,\\
\mathrm{PSF}_{\mathrm{tv}} &= \sum_{x,y} \bigl(\mathrm{PSF}(x+1,y) - \mathrm{PSF}(x,y)\bigr)^2
                           + \bigl(\mathrm{PSF}(x,y+1) - \mathrm{PSF}(x,y)\bigr)^2,\\
\mathrm{LP}_{\mathrm{tv}} &= \sum_{x,y} \bigl(\mathrm{LP}(x+1,y) - \mathrm{LP}(x,y)\bigr)^2
                          + \bigl(\mathrm{LP}(x,y+1) - \mathrm{LP}(x,y)\bigr)^2,\\
\mathrm{Loss} &= \alpha L_1 + \beta L_{\mathrm{MS\text{-}SSIM}} + \gamma\,\mathrm{Central}
                 + \delta\bigl(\mathrm{PSF}_{\mathrm{tv}} + \mathrm{LP}_{\mathrm{tv}}\bigr).
\end{align}

Where $\alpha = 0.875$, $\beta = 0.125$, $\gamma = 0.001$ and $\delta = 0.001$.

\subsection*{Training}

The training process for SIMFormer begins by loading a pre-trained model, leveraging the benefits of transfer learning since training a ViT from scratch on small datasets may not yield optimal performance. We use the pre-trained CLIP model~\cite{Radford2021CLIP} as the initialization parameter for SIMFormer. The initial training is conducted with a batch size of 96 over 70 epochs. The input patch size is set to $3\times16\times16$, and the mask ratio is 0.75. 

Next, a multi-stage training process is employed to gradually compress the rank of the light pattern representation, adjusting the low-rank coding rank from 512 to 32 (256, 128, 64, and finally 32). In each stage, the mask ratio is 0.25, with each stage running for 30 epochs and batch sizes adjusted to 64. After reducing the low-rank coding rank to 32, the model undergoes an extensive training stage with 2000 epochs, a batch size of 96, and a mask ratio of 0.25 for synthetic data / 0.75 for BioSR. See Supplementary Note 7 for details.

Throughout the process, various training parameters such as the patch size of $3\times16\times16$, crop size $80\times80$, PSF size $49\times49$, and learning rate of $10^{-4}$ are maintained. The raw images are linearly scaled up by a factor of 3, resulting in threefold enlarged SIM reconstruction images. The unscaled raw images are used when calculating the loss.

\subsection*{Self-distillation}

We use self-distillation to improve the performance of SIMFormer, a technique where a model is refined by training it to learn from its own predictions (Supplementary Note 8). Specifically, we first train SIMFormer using high signal-to-noise ratio (SNR) SIM images, allowing it to generate high-resolution predictions. These predictions are then used as soft labels to train a new instance of the model SIMFormer+ under conditions of lower photon counts and increased noise. All high signal-to-noise ratio versions of the images in the test set did not appear in the training set of SIMFormer and SIMFormer+. This approach allows SIMFormer+ to inherit the high-resolution capabilities of the original model, while significantly improving its robustness to noise.

\subsection*{Normalization}

Image reconstruction algorithms often produce results with ranges that differ from those of the ground truth images. To address this issue, we apply a commonly used normalization method to ensure a consistent data range. Specifically, percentile normalization is applied to both the ground truth and reconstructed images:
\[
\mathrm{Norm_p(image)} =
\frac{\mathrm{image} - \mathrm{Percentile}(\mathrm{image}, 0.1)}
{\mathrm{Percentile}(\mathrm{image}, 99.9) - \mathrm{Percentile}(\mathrm{image}, 0.1)}.
\]
In this equation, we use percentiles of the image to determine the normalization range. $\mathrm{Percentile}(\mathrm{image}, 0.1)$ represents the 0.1 percentile of the image, and $\mathrm{Percentile}(\mathrm{image}, 99.9)$ represents the 99.9 percentile of the image.

\subsection*{Decorrelation analysis}

The resolution is assessed through decorrelation analysis~\cite{Descloux2019Decorrelation}, which differs from Abbe’s principle by identifying the highest frequency through the examination of local peaks in the decorrelation functions. Details in Supplementary Note 6.

\subsection*{Computational Resource and Speed}

In this work, we use the JAX library~\cite{Bradbury2018JAX}, as detailed in Code File 1. JAX is a widely used open-source deep learning framework known for its efficiency and ease of use in building and deploying models. The model is trained on 8 NVIDIA Tesla A100 80~GB GPUs over a period of $7\times24$~hours. Inference with the trained model for a single SIM stack $(9, 512, 512)$ takes approximately 1~second.

\section*{Acknowledgments}

Fang-Wen Sun and Xiang-Dong Chen were supported by the National Natural Science Foundation of China (No.~62225506), CAS Project for Young Scientists in Basic Research (No.~YSBR-049), and the Innovation Program for Quantum Science and Technology (No.~2021ZD0303200). Xiang-Dong Chen was also supported by the Key Research and Development Program of Anhui Province (No.~2022b13020006). The authors would like to thank Drs. Yu Zheng, Yang Dong, Ce Feng and Shao-Chun Zhang for fruitful discussions.

\textbf{Author contributions.} Z.~W. conceived the project. F.~S. and T.~C. supervised the research. Z.~W., F.~S., and T.~C. designed the experiments. T.~W. and Z.~W. performed experiments. Z.~W., F.~S., and T.~C. wrote the manuscript. All authors discussed the results and commented on the manuscript.

\textbf{Competing interests.} The authors declare no conflict of interest.

\textbf{Data availability.} Data supporting the findings of this study are available from the corresponding author upon reasonable request.

\textbf{Use of large language models.} Portions of this manuscript were edited for grammar and stylistic consistency with the assistance of a large-language-model service. The tool was used only for language polishing; it did not generate research content, analysis, or conclusions. All edits were individually reviewed and approved by the authors, who remain fully responsible for the accuracy and integrity of the work.

\clearpage
\bibliographystyle{unsrt}
\bibliography{ref}
\clearpage

\section*{Figures}

\begin{figure}[H]
    \centering
    \includegraphics[width=\linewidth]{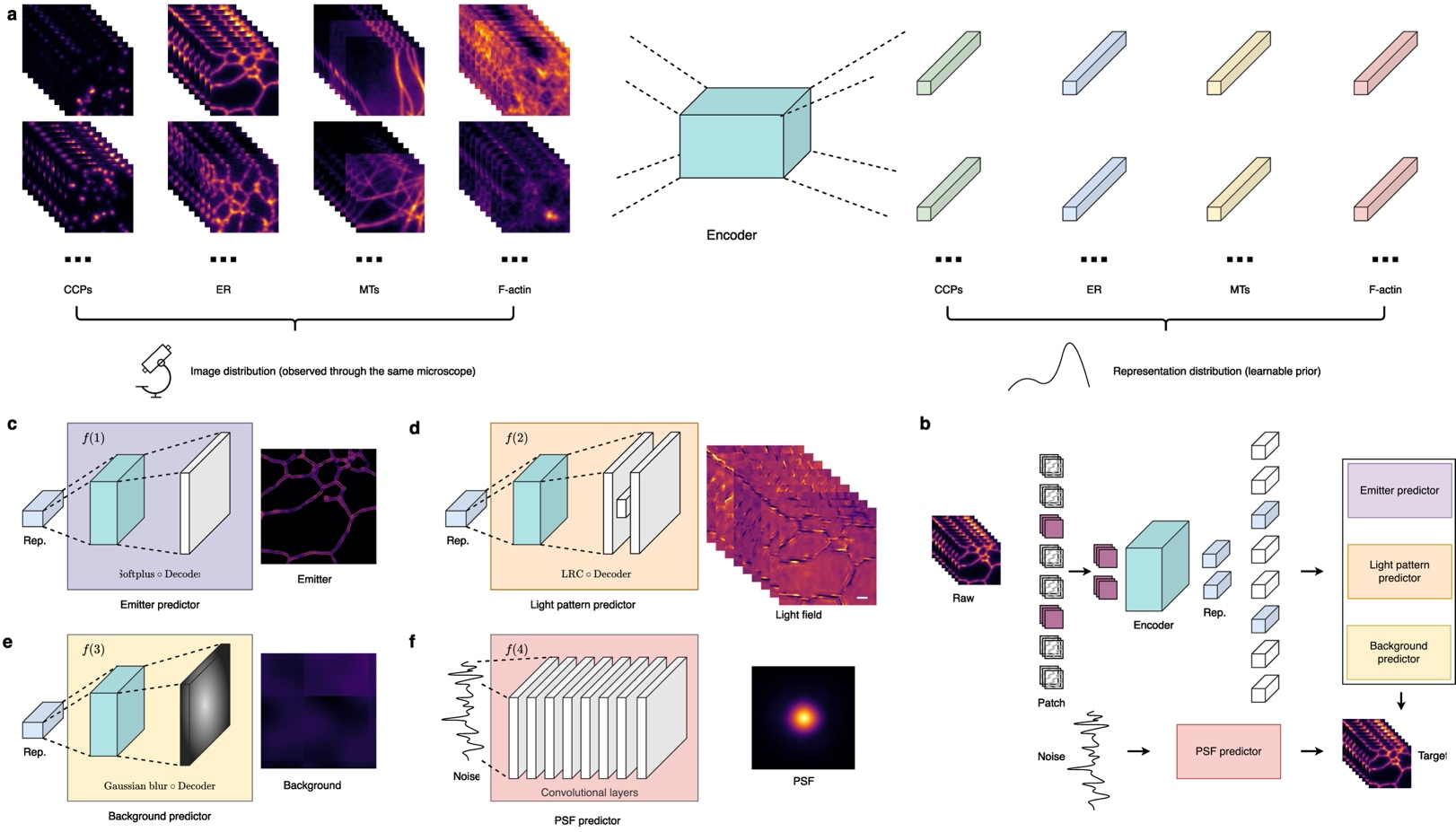}
    \caption{Fig.~1. Overview of SIMFormer. (A) The image encoder extracts representations from SIM raw images and is updated through self-supervised training. (B) SIMFormer is a self-supervised learning architecture based on masked autoencoders. (C) The representation is mapped to emitters using the decoder and Softplus activation. (D) The representation is mapped to light patterns using the decoder and low-rank coding (LRC). (E) The representation is mapped to the background using the decoder and Gaussian blur. (F) The PSF is generated from fixed noise using a convolutional network.}
    \label{fig:fig1}
\end{figure}

\begin{figure}[H]
    \centering
    \includegraphics[width=\linewidth]{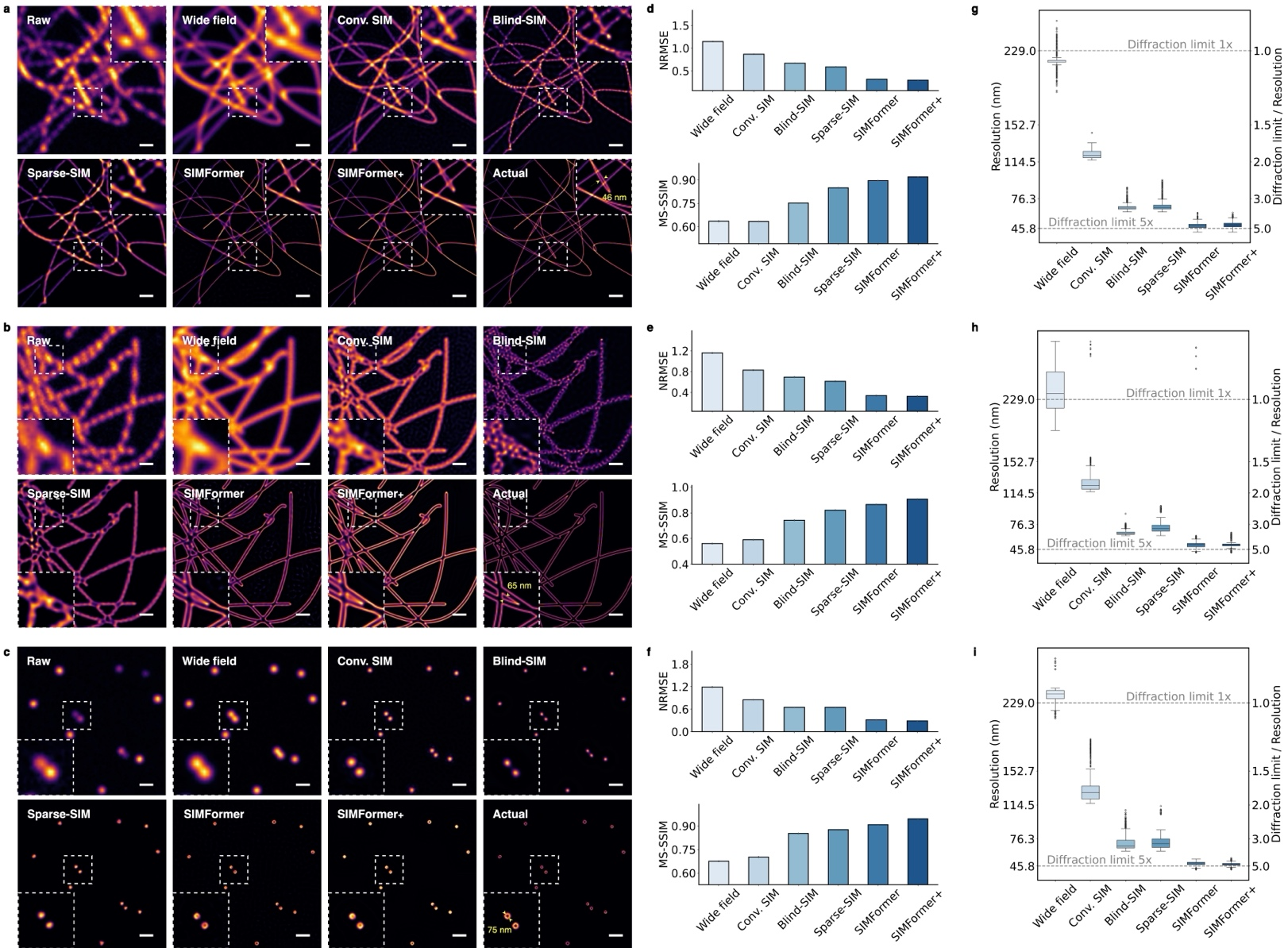}
    \caption{Fig.~2. Evaluation of estimated emitters for synthetic data. (A--C) Visual comparison of the super-resolution reconstruction of conventional SIM, Sparse-SIM, Blind-SIM, SIMFormer, and SIMFormer+ on the raw SIM images. (D--F) Quantitative analysis results using NRMSE and MS-SSIM. (G--I) Resolution obtained from the decorrelation analysis. (A) Visualization results of emitters with synthetic filament structures. (B) Visualization results of emitters with synthetic microtubule structures. (C) Visualization results for emitters with synthetic ring structures. Scale bar: 0.5~\textmu m. (D) NRMSE and MS-SSIM results of synthetic filament emitters. (E) NRMSE and MS-SSIM results of synthetic microtubular emitters. (F) NRMSE and MS-SSIM results of synthetic ring structure emitters. (G) Resolution of estimated emitters with synthetic filament structures. (H) Resolution of estimated emitters with synthetic microtubule structures. (I) Resolution of estimated emitters with synthetic ring structures. Box plots show the median (centre line), 25th--75th percentiles (box), and 1.5~$\times$~IQR whiskers; outliers are plotted as dots.}
    \label{fig:fig2}
\end{figure}

\begin{figure}[H]
    \centering
    \includegraphics[width=\linewidth]{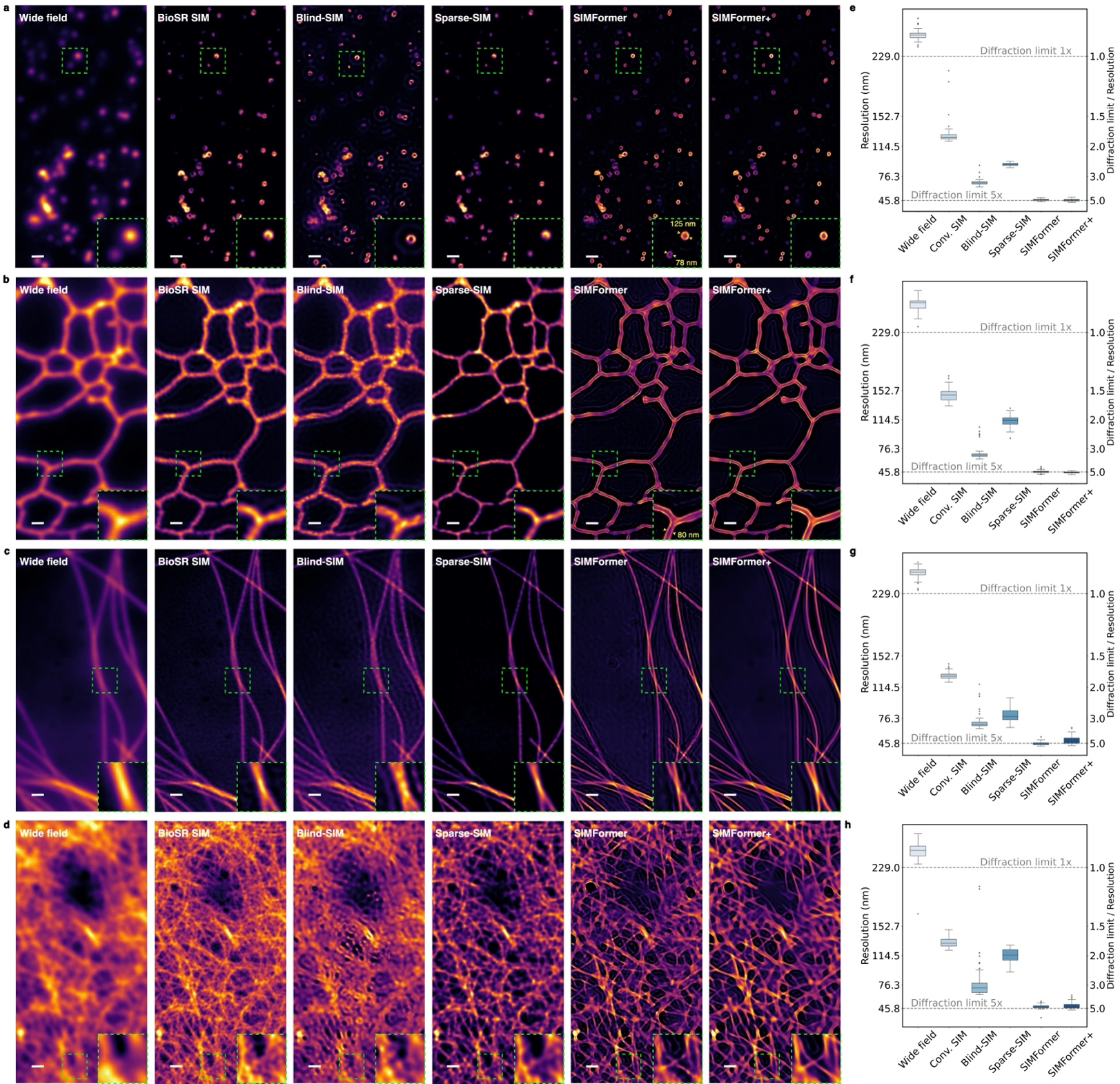}
    \caption{Fig.~3. Experiments on the BioSR dataset. (A--D) Visualization results of the BioSR data from wide field imaging, BioSR SIM, Blind-SIM, SIMFormer, and SIMFormer+. (E--H) Resolution obtained from decorrelation analysis. (A) Comparison results for ER. (B) Comparison results for CCPs. (C) Comparison results for MTs. (D) Comparison results for F-actin. Scale bar: 0.5~\textmu m. (E) Resolution for ER. (F) Resolution for CCPs. (G) Resolution for MTs. (H) Resolution for F-actin. Box plots show the median (centre line), 25th--75th percentiles (box), and 1.5~$\times$~IQR whiskers; outliers are plotted as dots.}
    \label{fig:fig3}
\end{figure}

\begin{figure}[H]
    \centering
    \includegraphics[width=\linewidth]{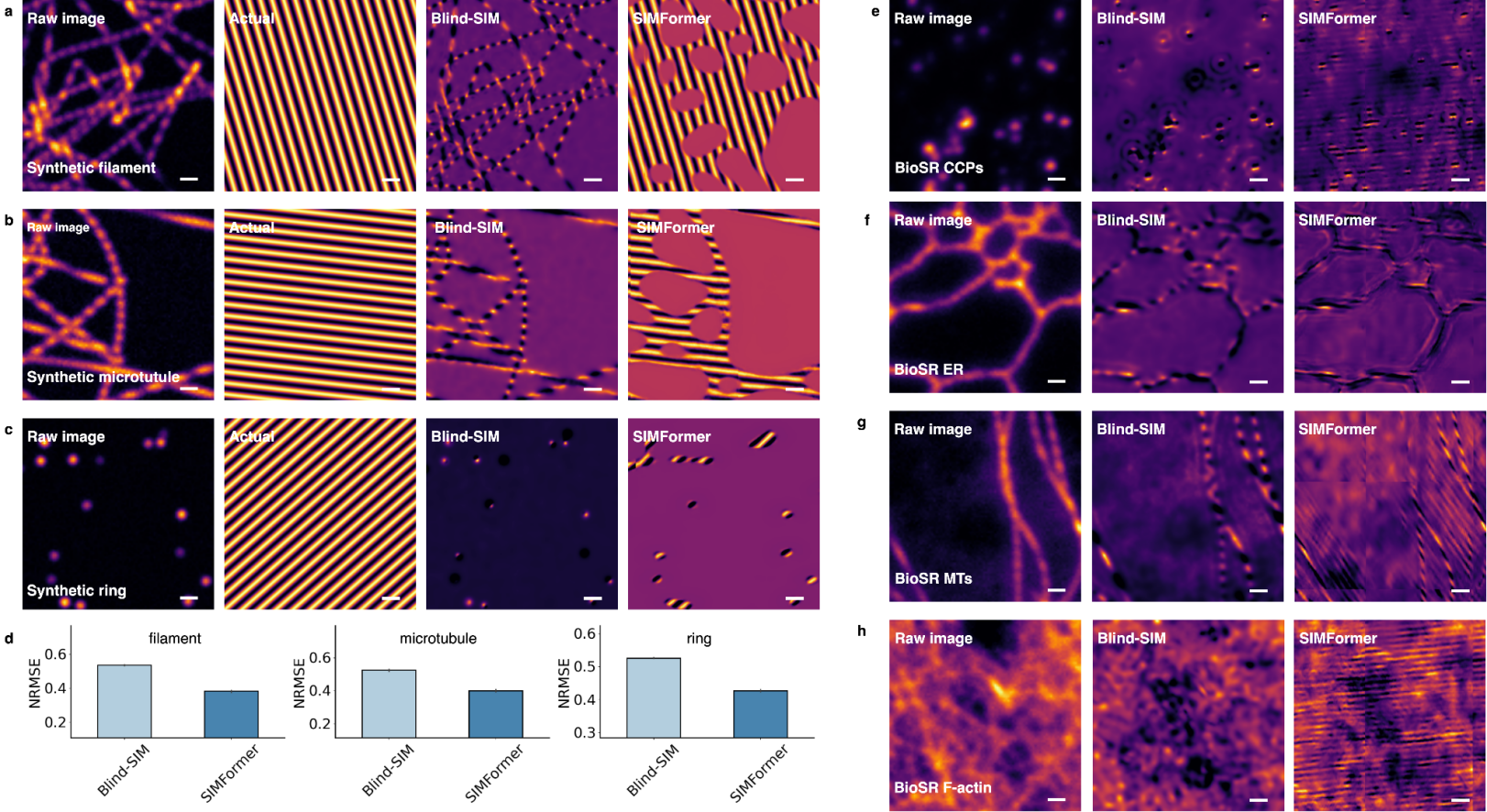}
    \caption{Fig.~4. Light pattern estimation. (A--C) Comparison of estimated light patterns from synthetic data. The emitters are (A) synthetic filament, (B) synthetic microtubule, and (C) synthetic ring. (D) Quantitative evaluation of the light pattern using NRMSE. (E--H) Comparison of estimated light patterns from BioSR data. The emitters are (E) CCPs, (F) ER, (G) MTs, and (H) F-actin.}
    \label{fig:fig4}
\end{figure}

\begin{figure}[H]
    \centering
    \includegraphics[width=\linewidth]{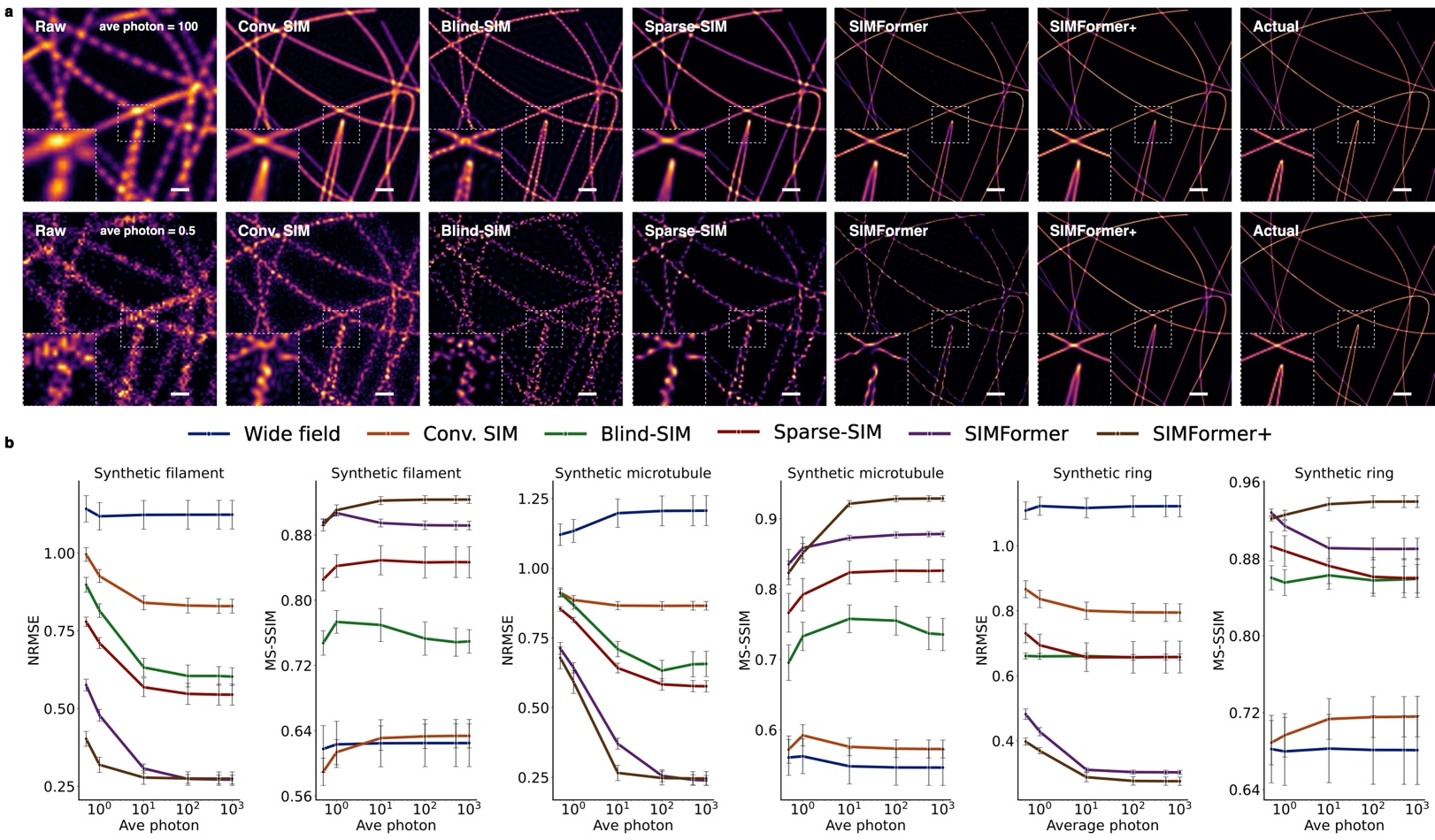}
    \caption{Fig.~5. Noise robustness of synthetic data. Comparison of the super-resolution reconstruction of conventional SIM, Sparse-SIM, Blind-SIM, SIMFormer, and SIMFormer+ on the raw SIM images with average photon count equal to 0.5, 1, 10, 100, 500, and 1000. (A) Comparative presentation of high and low photon-number results for synthetic filaments. (B) Quantitative evaluation using NRMSE and MS-SSIM. Scale bar: 0.5~\textmu m. Lines represent the mean; error bars denote the standard error of the mean.}
    \label{fig:fig5}
\end{figure}

\begin{figure}[H]
    \centering
    \includegraphics[width=\linewidth]{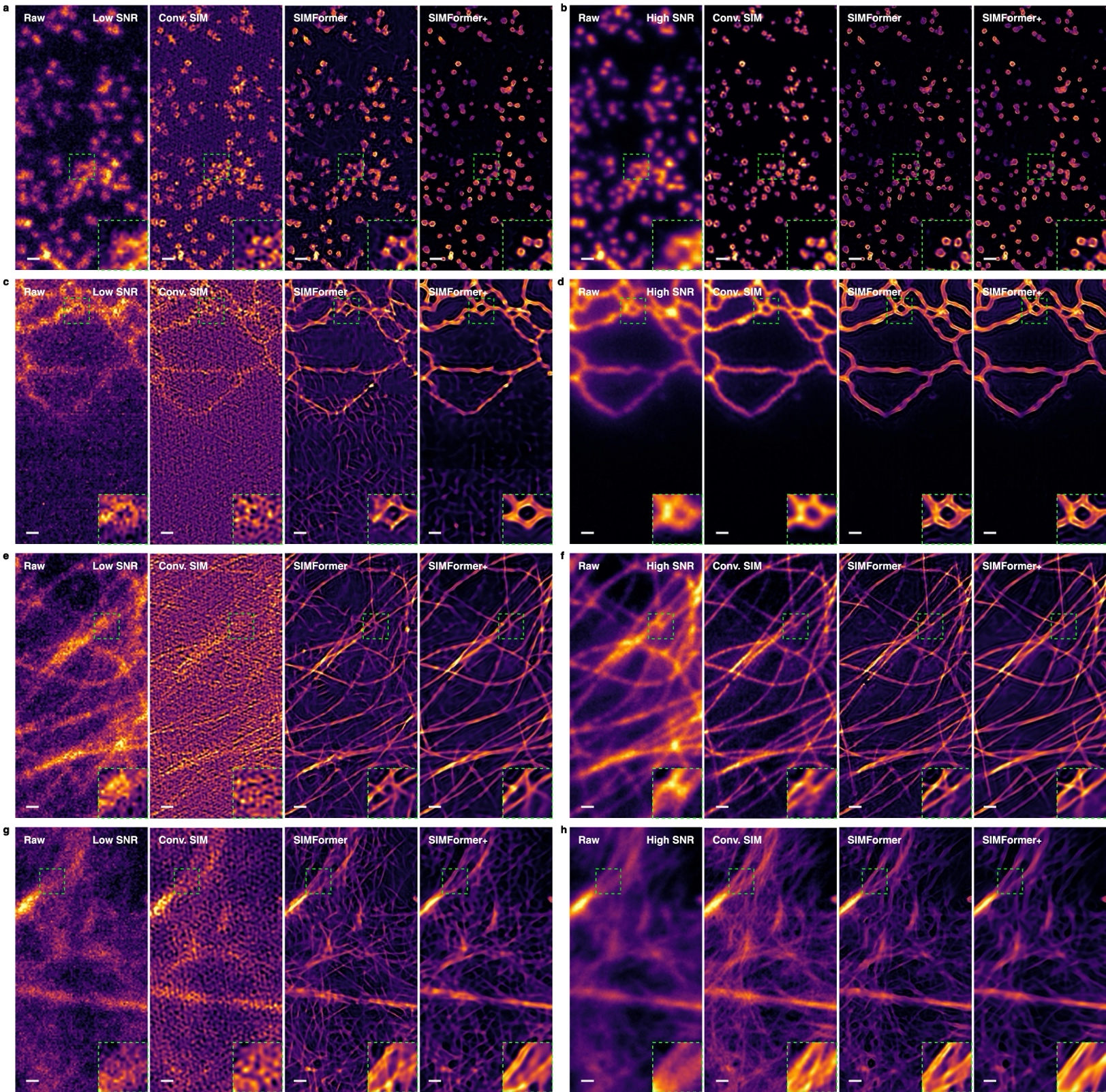}
    \caption{Fig.~6. Noise robustness on BioSR data. Comparison of SIM raw images, conventional SIM reconstruction, and SIMFormer+ with low signal-to-noise ratio (SNR) and high SNR. The raw images are (A) CCPs with low SNR, (B) CCPs with high SNR, (C) ER with low SNR, (D) ER with high SNR, (E) MTs with low SNR, (F) MTs with high SNR, (G) F-actin with low SNR, and (H) F-actin with high SNR. Scale bar: 0.5~\textmu m.}
    \label{fig:fig6}
\end{figure}

\clearpage
\section*{Extended Data Figures}

\begin{figure}[H]
    \centering
    \includegraphics[width=\linewidth]{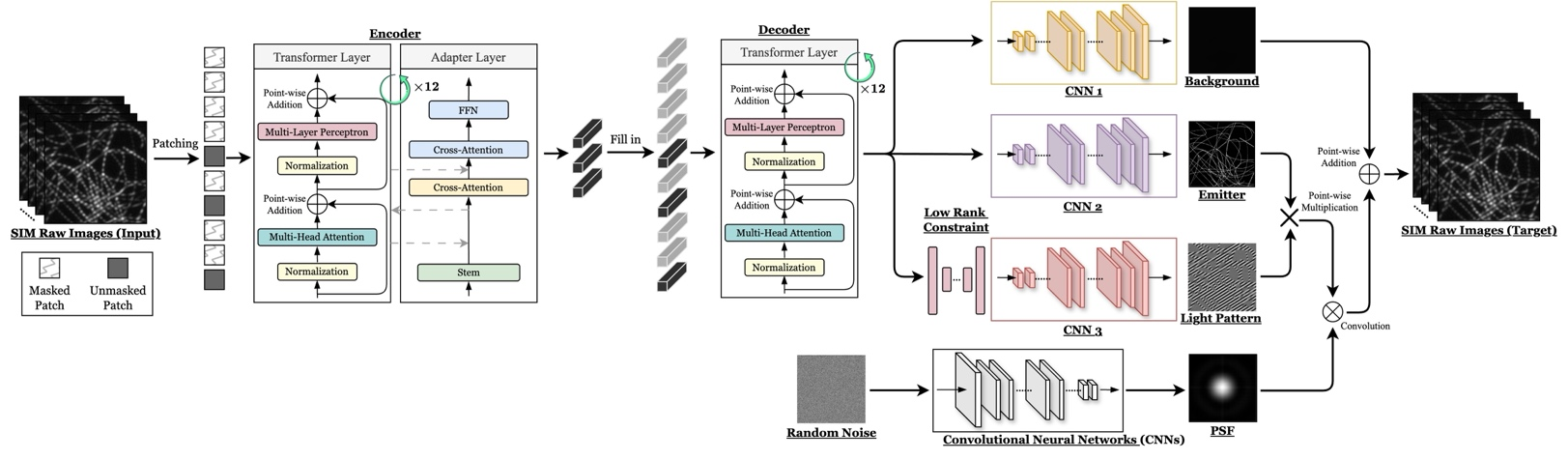}
    \caption{Extended Data Fig.~1. Network architecture of SIMFormer. The input is raw SIM images, and the output includes the background, emitters, light pattern, and PSF. The design uses a masked autoencoder with a transformer encoder, an adapter, and a transformer + CNN decoder.}
    \label{fig:extfig1}
\end{figure}

\begin{figure}[H]
    \centering
    \includegraphics[width=\linewidth]{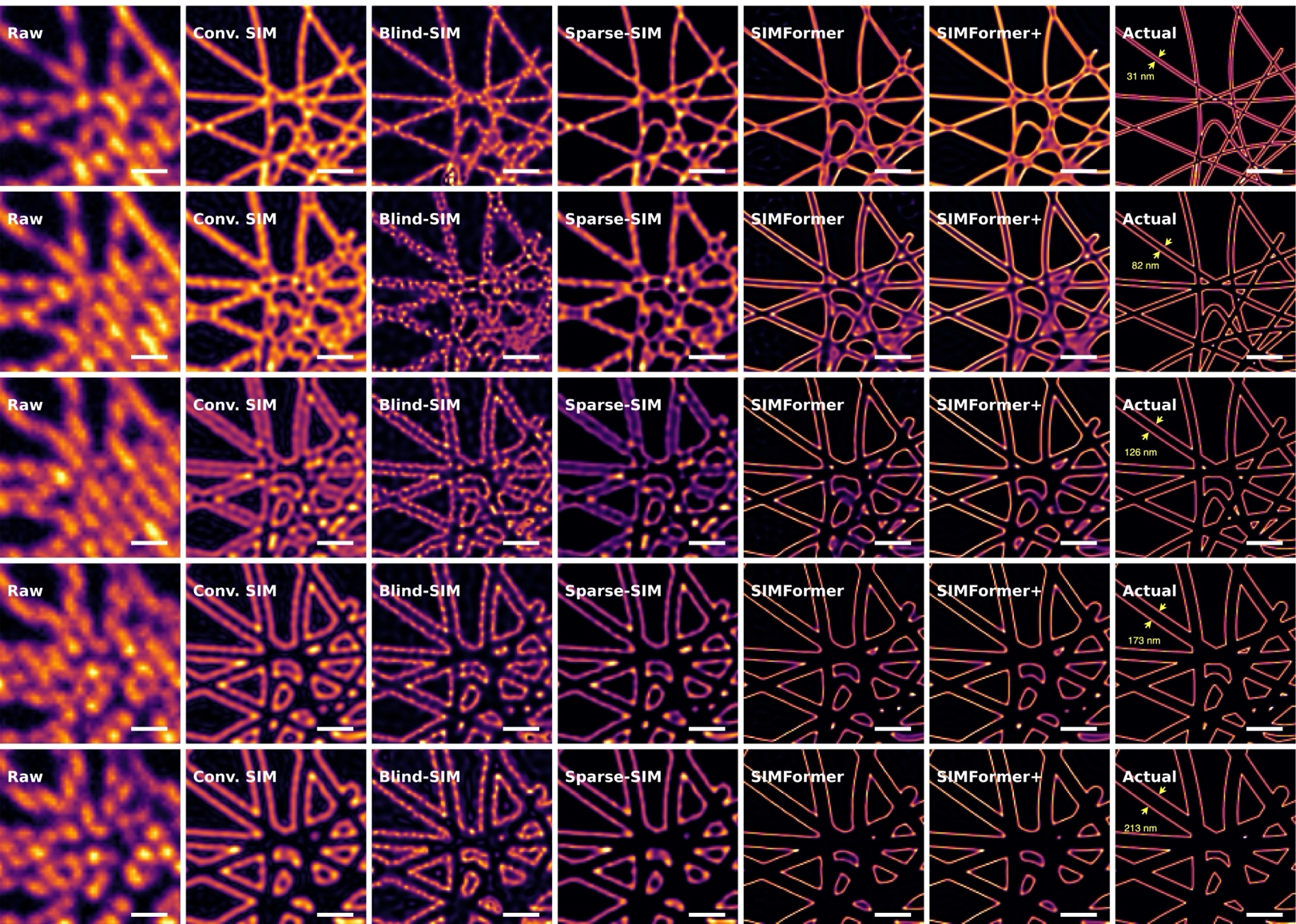}
    \caption{Extended Data Fig.~2. Representative synthetic microtubule structures results. Comparison of wide-field results, conventional SIM results, Blind-SIM results, Sparse-SIM results, SIMFormer results, and SIMFormer+ results. From top to bottom represents a gradual increase in microtubule width. Scale bar: 0.5~\textmu m.}
    \label{fig:extfig2}
\end{figure}

\begin{figure}[H]
    \centering
    \includegraphics[width=\linewidth]{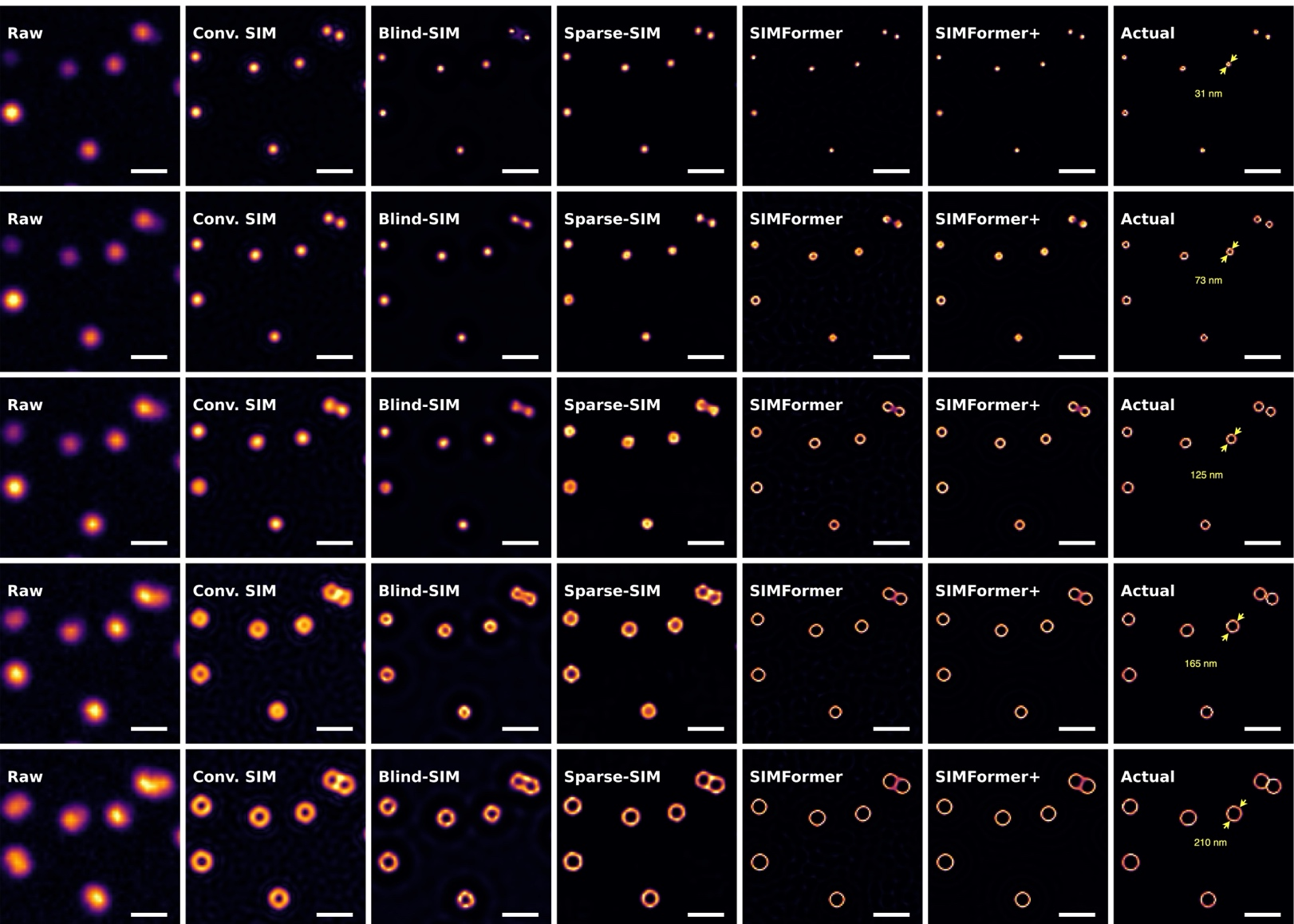}
    \caption{Extended Data Fig.~3. Representative synthetic ring structure results. Comparison of wide-field results, conventional SIM results, Blind-SIM results, Sparse-SIM results, SIMFormer results, and SIMFormer+ results. From top to bottom represents a gradual increase in the radius of the ring. Scale bar: 0.5~\textmu m.}
    \label{fig:extfig3}
\end{figure}

\begin{figure}[H]
    \centering
    \includegraphics[width=\linewidth]{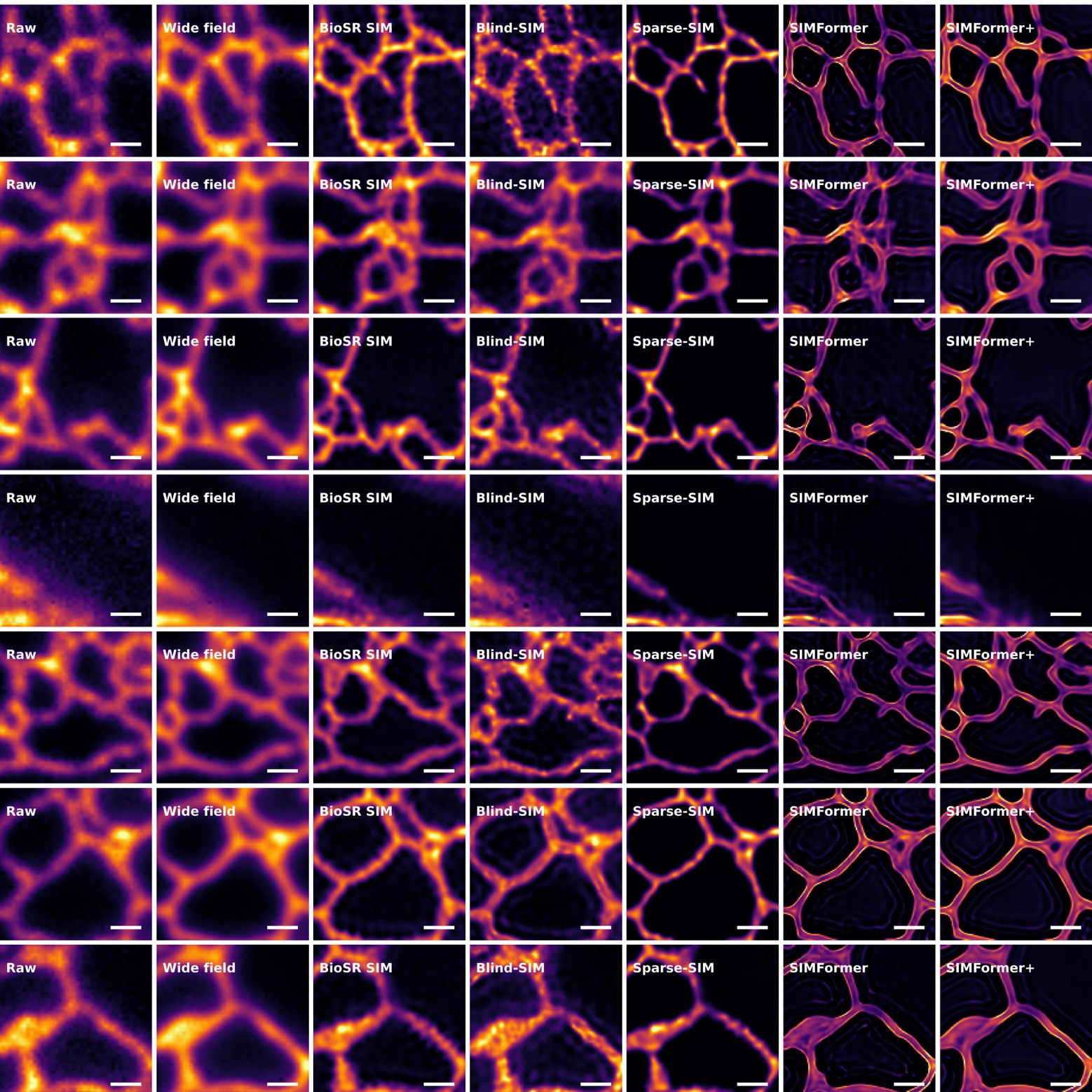}
    \caption{Extended Data Fig.~4. Representative ER results. Comparison of wide-field results, SIM results (BioSR), Blind-SIM results, Sparse-SIM results, SIMFormer results, and SIMFormer+ results. Scale bar: 0.5~\textmu m.}
    \label{fig:extfig4}
\end{figure}

\begin{figure}[H]
    \centering
    \includegraphics[width=\linewidth]{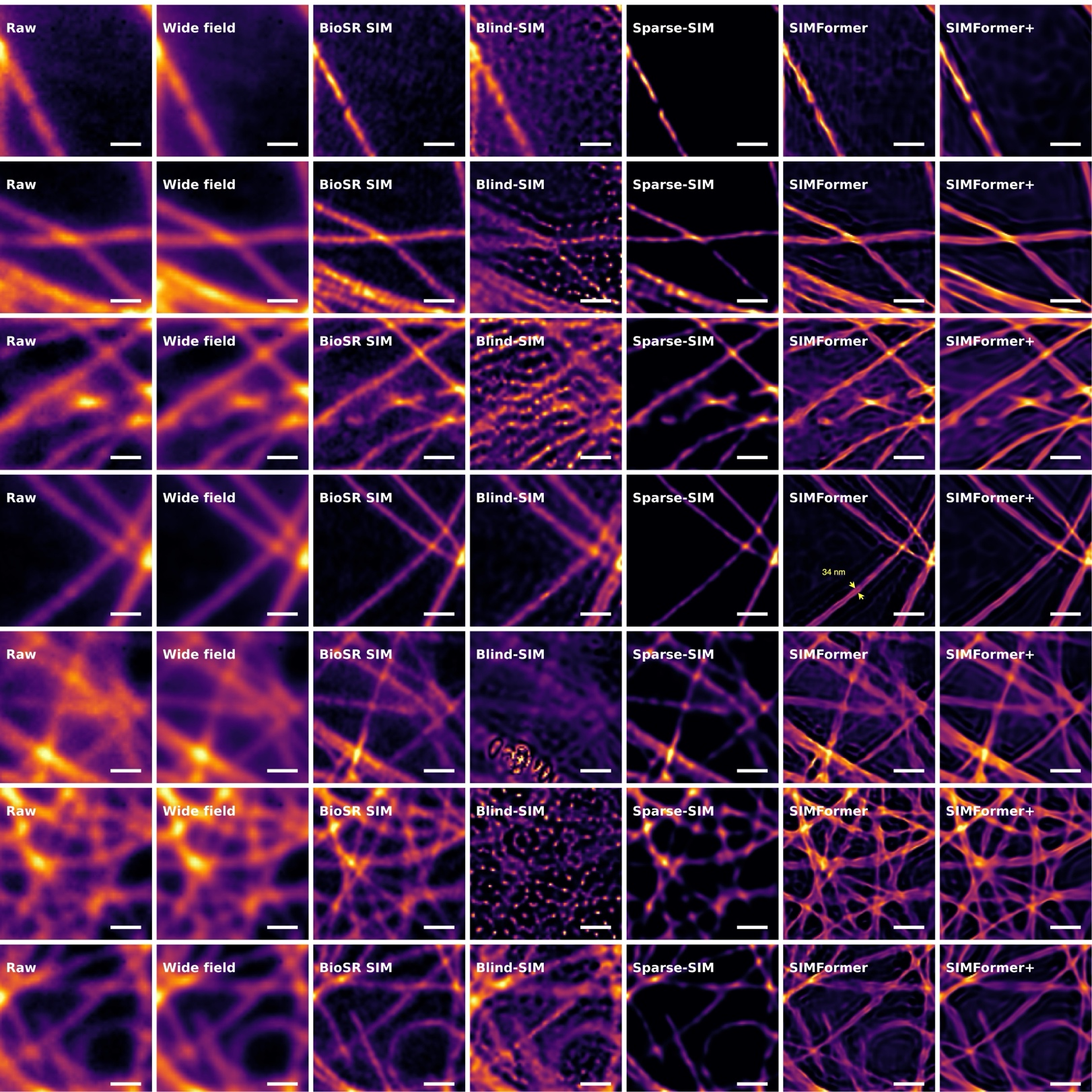}
    \caption{Extended Data Fig.~5. Representative MTs results. Comparison of wide-field results, SIM results (BioSR), Blind-SIM results, Sparse-SIM results, SIMFormer results, and SIMFormer+ results. Scale bar: 0.5~\textmu m.}
    \label{fig:extfig5}
\end{figure}

\begin{figure}[H]
    \centering
    \includegraphics[width=\linewidth]{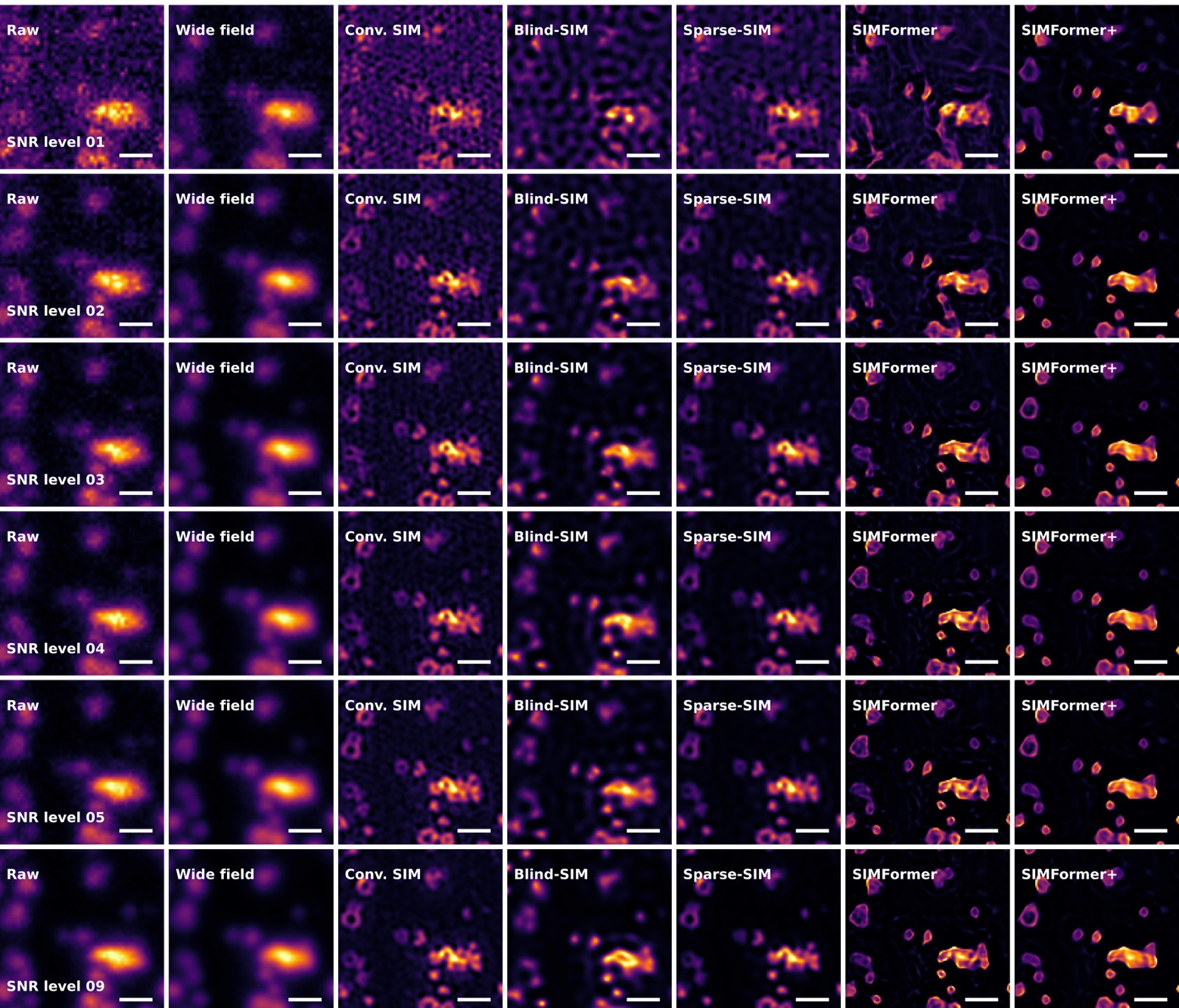}
    \caption{Extended Data Fig.~6. Noise robustness of CCPs data. Comparison of noise robustness among wide-field results, SIM results (fairSIM), Blind-SIM results, Sparse-SIM results, SIMFormer results, and SIMFormer+ results. Scale bar: 0.5~\textmu m.}
    \label{fig:extfig6}
\end{figure}

\begin{figure}[H]
    \centering
    \includegraphics[width=\linewidth]{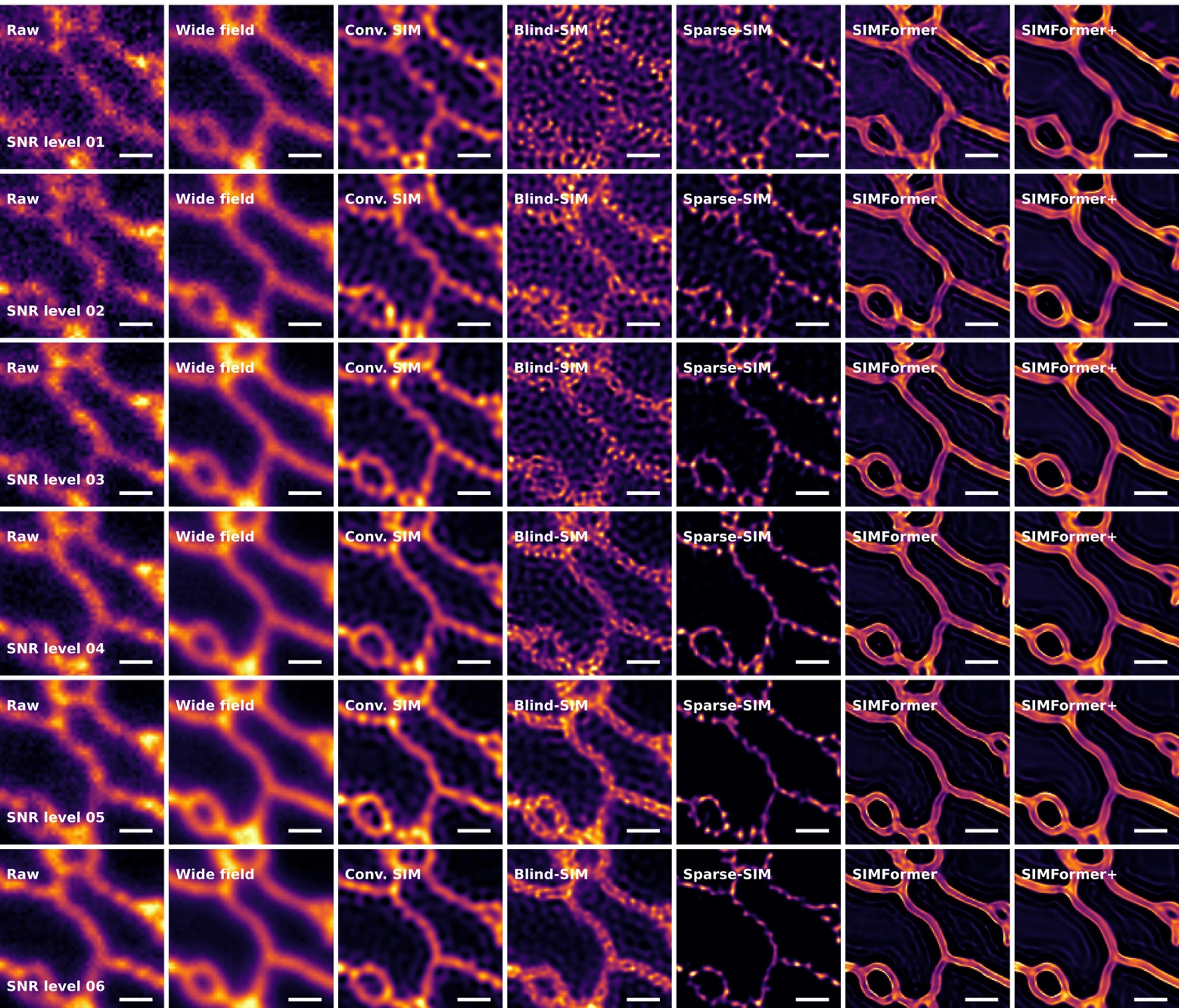}
    \caption{Extended Data Fig.~7. Noise robustness of ER data. Comparison of noise robustness among wide-field results, SIM results (fairSIM), Blind-SIM results, Sparse-SIM results, SIMFormer results, and SIMFormer+ results. Scale bar: 0.5~\textmu m.}
    \label{fig:extfig7}
\end{figure}

\begin{figure}[H]
    \centering
    \includegraphics[width=\linewidth]{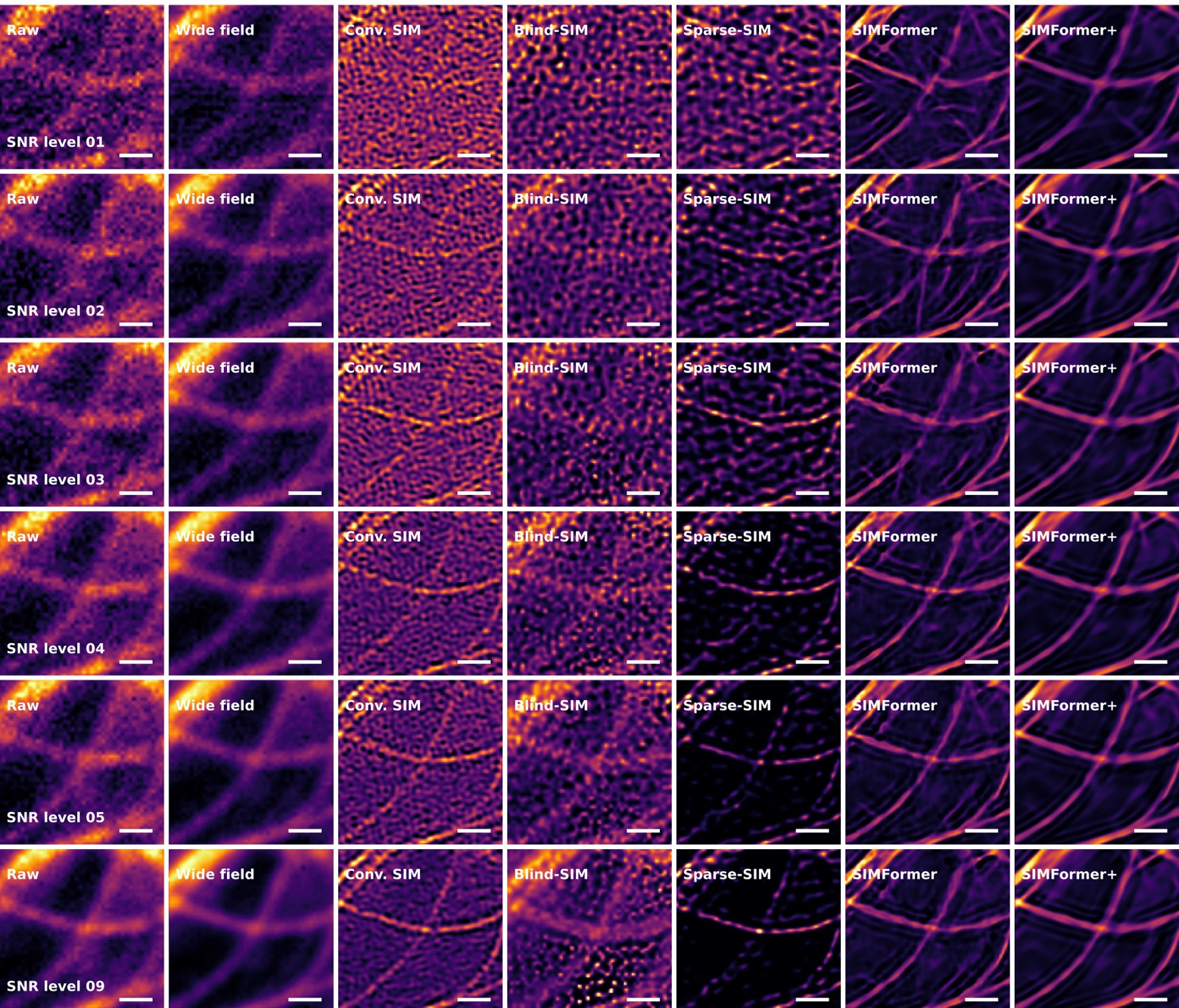}
    \caption{Extended Data Fig.~8. Noise robustness of MTs data. Comparison of noise robustness among wide-field results, SIM results (fairSIM), Blind-SIM results, Sparse-SIM results, SIMFormer results, and SIMFormer distilled results. Scale bar: 0.5~\textmu m.}
    \label{fig:extfig8}
\end{figure}

\begin{figure}[H]
    \centering
    \includegraphics[width=\linewidth]{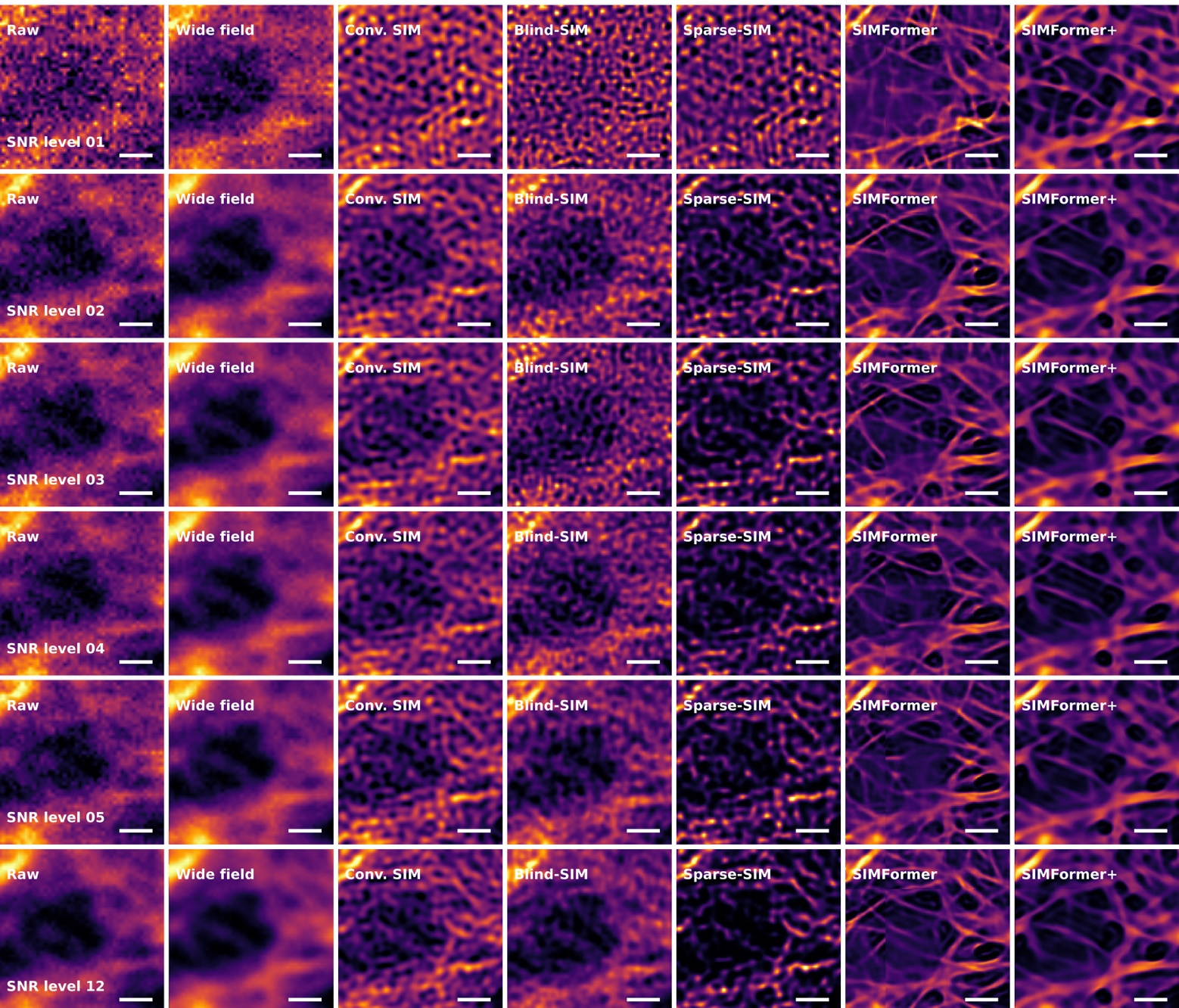}
    \caption{Extended Data Fig.~9. Noise robustness of F-actin data. Comparison of noise robustness among wide-field results, SIM results (fairSIM), Blind-SIM results, Sparse-SIM results, SIMFormer results, and SIMFormer distilled results. Scale bar: 0.5~\textmu m.}
    \label{fig:extfig9}
\end{figure}

\end{document}